\documentclass[prb,twocolumn,superscriptaddress]{revtex4}
\usepackage{times}
\usepackage{amsmath,amssymb,mathrsfs}
\usepackage[dvips]{graphicx}
\usepackage{hyperref}
\usepackage{paralist}
\usepackage{ifthen}
\usepackage{dsfont}
\usepackage{color}

\def\be{\begin{equation}}
\def\ee{\end{equation}}
\def\bea{\begin{eqnarray}}
\def\eea{\end{eqnarray}}

\begin{document}

\title{Symmetry protection of topological phases in one-dimensional quantum
spin systems}
\author{Frank Pollmann}
\affiliation{Max-Planck-Institut f\"ur Physik komplexer
Systeme, 01187 Dresden, Germany}
\author{Erez Berg}
\affiliation{Department of Physics, Harvard University, Cambridge,
Massachusetts 02138, USA}
\author{Ari M. Turner}
\affiliation{University of Amsterdam, 1090 GL Amsterdam, The Netherlands}
\author{Masaki Oshikawa}
\affiliation{Institute for Solid State Physics, University of Tokyo, Kashiwa 277-8581
Japan}
\date{\today}

\begin{abstract}
We discuss the characterization and stability of the Haldane phase in
 integer spin chains on the basis of simple, physical
arguments. We find that an odd-$S$ Haldane phase is a topologically non-trivial phase which is protected by
any one of the following three global symmetries: (i) the dihedral
group of $\pi$-rotations about $x,y$ and $z$ axes; (ii)
time-reversal symmetry $S^{x,y,z} \rightarrow - S^{x,y,z}$; (iii)
link inversion symmetry (reflection about a bond center),
consistently with previous results [Phys. Rev. B \textbf{81},
064439 (2010)]. On the other hand, an even-$S$ Haldane phase is
not topologically protected (i.e., it is indistinct from a trivial,
site-factorizable phase). We show some numerical evidence that
supports these claims, 
using concrete examples.
\end{abstract}

\maketitle

\section{Introduction}

States of matter can be classified into different phases, which are often
distinguished by (local) order parameters. Identification of phases
generally requires certain symmetries. For example, the ordered and
the disordered phases of the Ising model 
are sharply distinct 
only in the presence of the $Z_2$ symmetry of spin reversal. In
the absence of the symmetry, the two phases can be connected
without a phase transition and thus cannot be distinguished
uniquely. This phase transition corresponds to the spontaneous breaking of the $Z_2$
symmetry. Therefore, it is natural that the $Z_2$ symmetry is
required in this example to protect the ordered phase as a well-defined phase
distinct from the disordered phase.

Even if there is no symmetry which distinguishes the two phases, they
can still be separated by a transition. However, it is generically first order
and terminates at a critical end point. Thus, as in the case of
liquid/gas phases, there is a smooth path which connects the two phases,
without any phase transition. In this sense, 
in the absence of protection due to symmetry, phase transitions can
still exist but they do not generally
\emph{define} essentially distinct phases,

On the other hand, even when there is no local order parameter or
spontaneous breaking of a global symmetry, we sometimes find distinct
quantum phases separated by quantum phase transitions. We then attribute the
distinction to a non-trivial or ``topological phase''. While there are several known
characterizations of topological phases, the complete understanding in general dimensions is still lacking. The related question, what kind of
symmetry, if any, is required to protect the topological phase, is much less
obvious compared to the case of a standard spontaneous symmetry breaking.
In this article, we will consider
one-dimensional systems 
to establish
some intuition about this question. 

One of the simplest examples of a topological phase is the
Haldane phase in quantum spin chains.\cite{Haldane-1983,Haldane-1983a} As
predicted by Haldane, the Heisenberg antiferromagnetic (HAF) chain with an
integer spin $S$
\begin{equation}
\mathcal{H}_{\mathrm{HAF}} = J \sum_j \vec{S}_j \cdot \vec{S}_{j+1},
\label{HAFC}
\end{equation}
where $J>0$, has a nonzero excitation gap and exponentially decaying spin
correlation functions, while the same model is gapless and has power-law
correlations for a half-integer $S$.

Following Haldane's prediction, Affleck, Kennedy, Lieb, and Tasaki (AKLT)
presented model Hamiltonians, for which the ground state can be obtained
exactly.\cite{PhysRevLett.59.799,AKLT-CMP} In addition to providing a tractable model
in which the Haldane conjecture can be tested, the ground state (AKLT state)
was later found to exhibit several unexpected properties, such as a nonlocal
``string order'' and edge states, which extend also to states within the same phase.\cite{DenNijsRommelse}

On the other hand, despite the relative simplicity of quantum spin chains
and intensive study over several decades, the framework
for describing their topological properties has just recently been
understood.
 In fact, it was only recently that the
importance of inversion (parity) symmetry in the Haldane phase was pointed
out. Based on a field-theory (bosonization) analysis of a related boson
model, Berg \textit{et al.} pointed out in Ref.~\onlinecite{Berg-2008}  that the $%
S=1$ Haldane phase is distinct from other phases only in the
presence of inversion symmetry. Next, based on the Tensor
Entanglement Filtering Renormalization Group (TEFR) approach, Gu
and Wen stated that the $S=1$ Haldane phase is protected by the
\emph{combination} of the translation, complex conjugation (``time
reversal''), and inversion
symmetry.\cite{GuWen-TEFR2009,comment-TR} Gu and Wen pointed out
that the combined symmetry above protects the topological phase,
even when the existing characterizations (edge states and string
order) do not work. It turns out that the symmetry protection  can
be understood in terms of ``fractionalization'' of symmetry
operations at the edges and is reflected by non-trivial
degeneracies in the entanglement
spectrum.\cite{Pollmann-2010,Turner-2010} The fractionalization is
described precisely using projective representations of the
symmetry group. This approach was then generalized to any gapped
1D system and shown to give a complete procedure in one dimension
for identifying the topological phase of such systems.\cite{Chen-2011,Chen-2011a,Schuch-2010}%
Several 1D models in which symmetry fractionalization plays an important role have been 
studied recently, see for example Ref.~\onlinecite{Nakamura-2010,Hasebe-2011,Jiang-2010,Hirano-2008}.

In this paper, we illustrate the behavior of topological order in one dimension by
reexamining 
spin systems and the robustness of their topological phases on the basis of simple,
physical arguments and discuss a number of concrete examples. We say that the topological phase around the
AKLT state is robust if it cannot be adiabatically connected to
another, ``topologically trivial''
state, without going through a phase transition.
Here, ``topologically trivial'' means that the state is
site-factorizable, namely that the state is given by
a single tensor product of local states.
An example of such a topologically trivial state is
\begin{equation}
|\mathcal{D}\rangle =|0\rangle _{1}\dots |0\rangle _{L},
\label{eq:Dstate}
\end{equation}
which is the ground state of a chain with single-ion anisotropy
$D(S^z)^2$ in the limit of $D\rightarrow +\infty$.
(For a precise mathematical definition of the robustness
of the topological phases in one dimension, see also
Refs.~\onlinecite{Chen-2011,Chen-2011a}.) 
We show that for odd values of the spin $S$, our
results are consistent with
those of Refs.~\onlinecite{Berg-2008,Pollmann-2010}:
the AKLT state is robust as long as \emph{any one} of the three
symmetries mentioned in the abstract ($\pi$-rotation of the spin about $x$, $y$, $z$ axes, time reversal or inversion symmetry)  is respected. 
Surprisingly, these arguments suggest that other
systems, such as even-spin AKLT states and $S=1$ spin ladders with an even number
of legs, are \emph{not} topologically protected, even if all the symmetries
are respected. In particular, the $S=2$ AKLT phase is indistinct from a
trivial state, even if full SU$(2)$ symmetry is maintained. We show here
how to transform such states into one another,
giving numerical evidence that there are no phase transitions along the way.

This paper is organized as follows: We begin by discussing
the stability of the Haldane phase owing to
a hidden discrete symmetry in Section II,
with a clarification of the required symmetry. We then generalize
the Haldane phase in Section III to different symmetries and
discuss the concept of symmetry protected topological phases. In
Section IV we demonstrate concrete examples in the form of
matrix-product states and present numerical simulations to support
and illustrate our arguments.
In particular, we construct explicit paths which smoothly connect
the $S=2$ AKLT state to various site-factorizable states,
demonstrating that the former state is trivial. Our results are
summarized in Section V.

%

\section{Haldane phase in the presence of global D$_2$ symmetry}

First let us briefly discuss the hidden order and
edge states in the context of a hidden {$Z_2\times Z_2$}{\ } symmetry.
Although this concept had been developed in early 1990's,
the symmetry of the Hamiltonian required for this mechanism
has not been discussed explicitly.
Here we also clarify the required symmetry, which could be
understood as one of the symmetries protecting the Haldane
phase as a distinct, topological phase.

It is believed
that the ground state of the standard Heisenberg chain belongs to the Haldane
phase, which also includes the translationally invariant
Affleck-Kennedy-Lieb-Tasaki (AKLT) state. The $S=1$ AKLT state exhibits the
following two remarkable properties: \textbf{(I)} Free $S=\frac12$ degree of
freedom appearing at each end of the chain in the case of open boundary
conditions. Namely, the ground state of the AKLT Hamiltonian is 4-fold
degenerate due to the $2^2$ edge states, although the ground state is unique
in the case of periodic boundary conditions; \textbf{(II)} A nonlocal order
measured by the string order parameter~\cite{DenNijsRommelse}
\begin{equation}
\mathcal{O}_{\mathrm{str}}^{\alpha} \equiv \lim_{|j-k| \rightarrow \infty}
\langle S^\alpha_j e^{i \pi \sum_{j \leq l < k} S^\alpha_l} S^\alpha_k
\rangle .
\end{equation}
These two features turned out to be characteristics of
not only the AKLT state, but rather of the $S=1$ ``Haldane phase'',
which includes the ground states of the AKLT model and
the $S=1$ antiferromagnetic Heisenberg chain.
In fact, the degeneracy due to the edge states is split for
a generic open chain in the Haldane phase, with a finite length.
However, the splitting is exponentially small for
longer chains, resulting in $4$-fold quasi-degenerate
ground states below the Haldane gap.~\cite{Kennedy1990}
Numerical calculations have shown that
the string order parameter is also nonvanishing
within the Haldane phase.

Kennedy and Tasaki~\cite{KennedyTasaki-PRB1992} unified these two
apparently unrelated features as
consequences of hidden symmetry breaking. This concept is introduced as follows.
We introduce a nonlocal unitary transformation defined by
(see also Ref.~\onlinecite{MO-HiddenSym1992})
\begin{equation}
{U_{\mathrm{KT}}} = \prod_{j<k}\exp{\left( i\pi S^z_j S^x_k \right)}.
\label{UKT}
\end{equation}
This transforms spin operators as
\begin{align}
U_{\mathrm{KT}} S^x_j U_{\mathrm{KT}}^{-1} &=
S^x_j \exp{(i \pi \sum_{k>j} S^x_k)}, \\
U_{\mathrm{KT}} S^y_j U_{\mathrm{KT}}^{-1} &=
\exp{(i \pi \sum_{k<j} S^z_k)}
S^y_j \exp{(i \pi \sum_{k>j} S^x_k)}, \\
U_{\mathrm{KT}} S^z_j U_{\mathrm{KT}}^{-1} &=
\exp{(i \pi \sum_{k<j} S^z_k)}
S^z_j.
\end{align}
Although these are nonlocal operators with ``strings'',
the Heisenberg chain Hamiltonian~\eqref{HAFC} is transformed into
a Hamiltonian with only short-range interactions:
\begin{eqnarray}
\tilde{H} =\sum_j &\Big(& S_j^x\exp(i\pi S^z_{j+1})S^x_{j+1} \nonumber\\
&+& S_j^y\exp(i\pi (S^z_{j} + S^x_{j+1}))S^y_{j+1} \nonumber\\ 
&+& S_j^z\exp(i\pi S^z_{j})S^z_{j+1}\Big).
\end{eqnarray}
This is thanks to a cancellation of string factors
similar (although not identical) to that in Jordan-Wigner
transformation.

The transformation~\eqref{UKT}
can be applied to a wide class of spin chain Hamiltonians.
For the transformation to be useful, the transformed Hamiltonian
must have only short-range interactions.
We point out that the sufficient and necessary
condition the Hamiltonian must satisfy for the transformed Hamiltonian to have local interactions is
that it has to have a global discrete symmetry
with respect to rotation by angle $\pi$ about $x$, $y$, and $z$
axes (i.e., $\prod_j \exp(i\pi S_j^x)$, and similar for y and z).
This symmetry group, sometimes
called the dihedral group D$_2$, is equivalent to {$Z_2 \times Z_2$}, since
the product of $\pi$-rotations about the $x$ and $z$ axes gives the $\pi$%
-rotation about the $y$ axis.
We note that, although global D$_{2}$ invariant models and time-reversal
invariant ones have a large overlap, they are not identical. For example,
the anisotropic perturbation $%
\sum_{j}(S_{j}^{z}S_{j+1}^{z}+S_{j}^{z}S_{j+2}^{x})$ is time-reversal
invariant but not D$_{2}$ invariant. On the other hand, $%
\sum_{j}S_{j}^{x}S_{j+1}^{y}S_{j+2}^{z}$ is D$_{2}$ invariant but not
time-reversal invariant.

The transformed \emph{Hamiltonian}
$U_{\mathrm{KT}} \mathcal{H} U_{\mathrm{KT}}^{-1}$ has the same $Z_2
\times Z_2$ symmetry as the original one. This is because
the $\pi$ rotations around $x,y,$ and  $z$ 
transform into themselves under $U_{\mathrm{KT}}$ 
(e.g., $e^{i\pi S_l^z}$ commutes with each of the
factors $e^{i\pi S_{z}^jS_x^k}$) 
However, as the states of the spins are transformed
in a nonlocal way, a state without any broken global symmetry
may be transformed into a state
with long-range 
ferromagnetic order; that is, there may be 
\emph{hidden} symmetry breaking.

\begin{figure}[tbp]
\includegraphics[width=5cm]{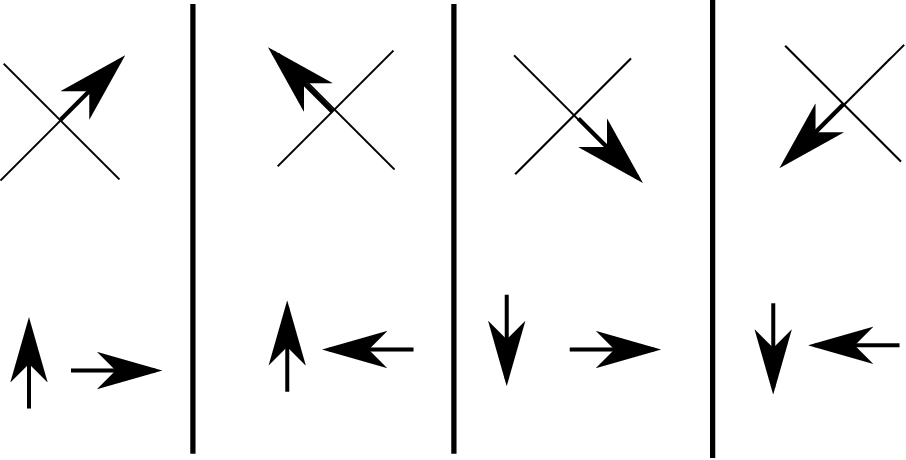}
\caption{Four symmetry broken states (upper panel) which are obtained by applying the non-local transformation $U_{KT}$ to the  degenerate edge states (lower panel). Note that the arrows in the upper panel represent the spin polarization in the bulk, while in the lower panel they represent the spin polarization at the edges.}
\label{fig:railwaycrossing}
\end{figure}

The symmetry breaking in the transformed system is an indication
of edge states in the original Hamiltonian. Note
that edge states on a finite chain can break the $Z_2\times Z_2$
symmetry. 
Although this symmetry breaking occurs only at the ends,
the non-local transformation spreads this symmetry breaking through the entire bulk. 
In fact, the $Z_2\times Z_2$ symmetry is broken completely
in the bulk, implying a 4-fold degenerate
set of ground states with magnetization along diagonal
directions as illustrated in Fig.~\ref{fig:railwaycrossing}. The $z$ and $x$ components of the magnetization in the bulk after the non-local transformation determine
the $z$ component of the spin-$\frac{1}{2}$ at the left end and the $x$-component
of the spin at the right end in the original system, respectively. The string order 
of the original system\eqref{HAFC} is also simple
to understand in terms of the hidden breaking of the symmetry:
it is the result of applying the 
 Kennedy-Tasaki transformation~\eqref{UKT} 
 to the usual ferromagnetic order parameter.

To understand the correspondence between the edge states and the broken
symmetry in the bulk, it is easiest to consider the AKLT state.
 The
four  degenerate  states  which transform into the four symmetry-broken states
are defined by giving the $z$ component of the free spin-$\frac{1}{2}$ at the
left end and the $x$ component of the spin-$\frac{1}{2}$ at the right end
definite values. 
To see this, start by fixing just the spin at the
left end to $S_{z,\mbox{\scriptsize left}}=+\frac{1}{2}$. Then expand this state
in terms of $S_z$ eigenstates.  
The string order is perfect in the AKLT state, meaning that in every component of the wave function, if we erase the sites with $S_z=0$, we get a chain with a perfect antiferromagnetic order, $\vert ... -1, +1, -1, +1, .... \rangle$.\cite{PhysRevLett.59.799,AKLT-CMP}  Thus the first non-zero spin must be $+1$ in each term and from then on the non-zero $S_z$'s
alternate between $\pm1$.  When $U_{KT}$ is applied, it flips the direction of every second nonzero
spin, so that all the sites
end up in either the $|S_z=+1\rangle $ or $|S_z=0\rangle$ state.
When the $x$-component of the spin-$\frac{1}{2}$ at the right end is also
fixed, one can likewise argue that each site is in the $|S_x=+1\rangle$
state or the $|S_x=0\rangle$ state (working in the $S_x$ basis instead).
But these two conditions
together uniquely determine the state of every site. Note that the only spinor that has only $+1$ and $0$ states in
both the x and z basis is
$\sqrt\frac{2}{3}|S_x=1\rangle +\sqrt\frac{1}{3}|S_x=0\rangle=
i\left(\sqrt\frac{2}{3}|S_z=1\rangle +\sqrt\frac{1}{3}|S_z=0\rangle\right)$.
So after applying the \emph{non-local} transformation $U_{KT}$, the wave function is just a product state with this state on every site, spontaneously breaking the symmetry all along the chain. 

Similar arguments can be applied to relate any of the four polarizations of the edge states to the four broken symmetry states after the non-local transformation, as shown in  Fig.~\ref{fig:railwaycrossing}.
A closely related analysis from a different perspective was recently
discussed in Ref.~\onlinecite{Okunishi-disentangler}.


The stability of the Haldane phase for $S>1$ has been less
understood. Once the transformation is written as Eq.~\eqref{UKT}, it can be
readily applied to any integer $S$, and the transformation of 
the Hamiltonian and 
the string order parameter remain the same. However, it turns out
that the hidden {{$Z_2 \times Z_2$}{\ }} symmetry is spontaneously
broken in the translationally invariant AKLT state only if $S$ is
odd, but unbroken if $S$ is even.~\cite{MO-HiddenSym1992} This can
also be seen by counting the degeneracy of the edge states.
Therefore, with regard to the hidden {$Z_2 \times Z_2${\ }}
symmetry, the even-$S$ AKLT states are indistinguishable from a
trivial disordered state. However, the physical meaning of this
finding was not well understood; it was unclear if the even-$S$
AKLT states are really indistinguishable from a trivial state, or
whether they are 
distinct from a trivial state by another, unknown criterion.

\section{Stability of the Haldane phase}

In the following, we discuss different ways to understand topological phases
in one-dimensional quantum spin systems without referring to the
hidden {$Z_2 \times Z_2$} symmetry.
We find that the odd-$S$ and even-$S$ AKLT state differ in the
robustness of the topological phase, as was suggested, in retrospect, by the
hidden {$Z_2 \times Z_2$}\ symmetry analysis.

\subsection{Characterization by edge states in the presence ot Time reversal symmetry}

\label{sec:edge} Let us now discuss the topological phase, from the
viewpoint of ``edge physics''. Here we apply the idea similar to what was
used to characterize the quantum spin Hall insulator.~\cite%
{PhysRevLett.95.146802} As long as the gap does not close in the bulk, we
may focus on the nearly degenerate ground states corresponding to the edge
states. The spin-$S$ AKLT state with open boundary conditions has a spin-$%
S/2 $ edge degree of freedom at each end, and thus $(S+1)$-fold degeneracy
at each end. In general, if we introduce a perturbation to the Hamiltonian,
the edge degeneracy is expected to be lifted. However, if the edge spin is
half-integer, namely for the odd-$S$ AKLT state, as long as the Hamiltonian
has time-reversal symmetry, the two-fold Kramers degeneracy at each edge
should remain. As a consequence, the odd-$S$ AKLT state must be separated
from a trivial disordered state by a quantum phase transition. In this
sense, the topological phase in the odd-$S$ AKLT state is robust and
protected by time-reversal symmetry.

On the other hand, for the even-$S$ AKLT state, the edge spin is an integer.
Thus the degeneracy is lifted by a generic perturbation even if the
Hamiltonian is invariant under time reversal, because there is no Kramers
degeneracy. If the Hamiltonian is SU$(2)$ invariant, the lowest $S+1$
degeneracy related to the edge should remain up to a finite strength of the
perturbation. However, as a function of the perturbation strength, there is
always a possibility that a $S=0$ state separates from the bulk, crosses the
$S+1$ multiplet and becomes the ground state. Note that in such a process,
the bulk gap need not close anywhere. Thus, it seems that an even-$S$ AKLT
state is, strictly speaking, \emph{indistinct} from a trivial state, regardless of the presence of time reversal or SU$(2)$
symmetries. We demonstrate this explicitly in Sec.~\ref{example}, where we show that
the $S=2$ AKLT state can be smoothly connected to a fully dimerized state.



\subsection{Inversion symmetry of a ring}

\label{sec:inversion}
\begin{figure}[tbp]
\includegraphics[width=3.3cm]{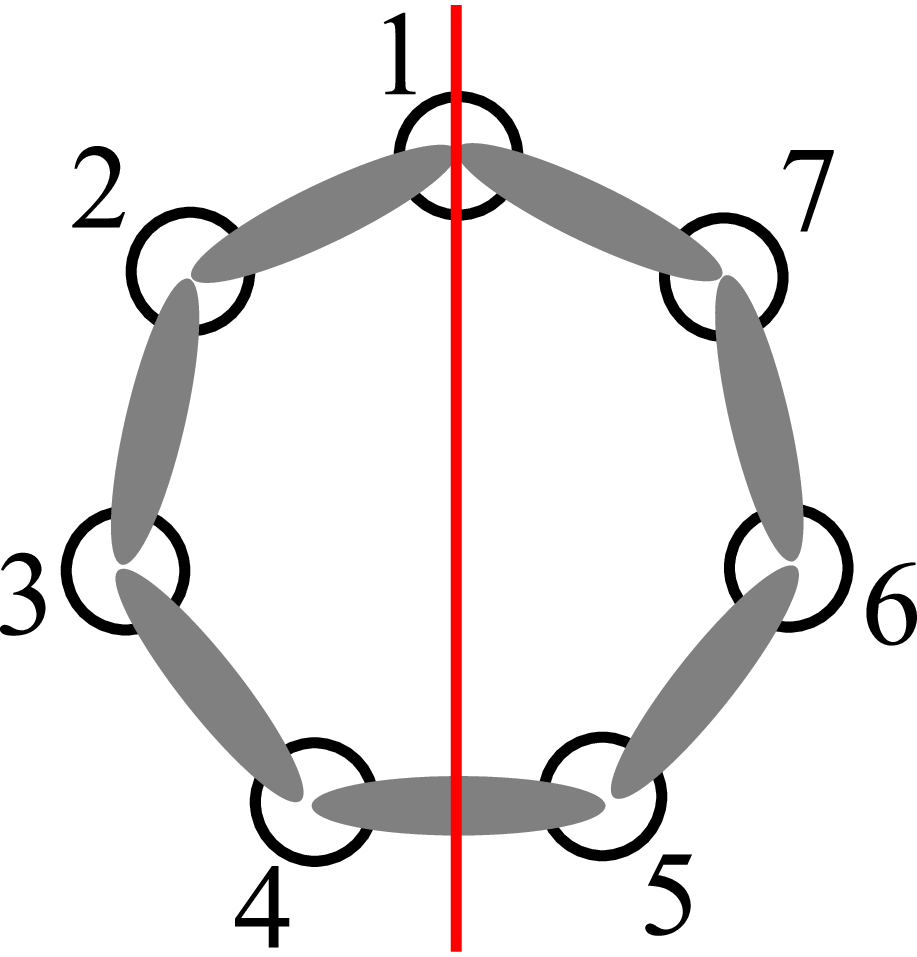}
\caption{(Color online) The $S=1$ AKLT state on a ring with $L=7$ sites. The connecting
lines represent spin-$\frac12$ singlets. We consider the lattice inversion about
the vertical line.}
\label{fig:AKLTringparity}
\end{figure}

We will now argue that the odd-$S$ Haldane phase is
also protected just by link inversion symmetry (lattice inversion about the
center of a bond). To illustrate the point, it is convenient to consider
first an AKLT state on a chain of \emph{odd} length $L$ with periodic
boundary conditions. Although the system is frustrated for odd $L$, the
ground state of the AKLT model is still unique, reflecting the short-range
spin correlations. For example, we discuss the $S=1$ AKLT state for $L=7$ as
shown in Fig.~\ref{fig:AKLTringparity}, and inversion $\mathcal{I}$ about
the vertical line.

Let us recall the original AKLT construction, starting from two $S=\frac12$'s
per site, and denote a valence bond (singlet of two $S=\frac12$'s) between sites
$j$ and $k$ by $|(j,k)\rangle$. The valence bond $|(j,k)\rangle$ is
antisymmetric under inversion, namely the exchange of $j$ and $k$. Thus,
under inversion $\mathcal{I}$, the valence bond $|(4,5)\rangle$ crossing the
line changes sign: $\mathcal{I} |(4,5)\rangle = - |(4,5)\rangle$. The
other valence bonds are flipped as well, for example $\mathcal{I} |(3,4)
\rangle = - |(5,6) \rangle$. However, being paired with $\mathcal{I}%
|(5,6)\rangle = -|(3,4)\rangle$, we find $\mathcal{I} |(3,4)\rangle|(5,6)%
\rangle = |(3,4)\rangle |(5,6)\rangle $. The symmetrization operation in the
AKLT construction is also invariant under $\mathcal{I}$. Thus we obtain
\begin{equation}
\mathcal{I} | \Psi^{\mbox{\scriptsize AKLT}}_{S=1} \rangle_{L=7} = -| \Psi^{%
\mbox{\scriptsize AKLT}}_{S=1} \rangle_{L=7} .
\end{equation}
(For a related discussion in a different setting, see Ref.~%
\onlinecite{JPSJ.66.3944}) The same argument can be easily applied to
higher-spin AKLT states: an odd-$S$ AKLT state is odd under inversion
on a ring with any odd length $L$ because there are
an odd number of valence bonds on every link. Even if we introduce perturbations to the
odd-$S$ AKLT model, the ground state on the odd length ring should still be
odd under inversion, as long as the Hamiltonian respects inversion symmetry
and the gap does not close. On the other hand, a trivial state given by a
tensor product of local states, such as $|\mathcal{D}\rangle$ defined in
eq.~\ref{eq:Dstate}, is even under
$\mathcal{I}$. Therefore we conclude that there must be a phase transition
between the odd-$S$ AKLT state and the trivial $|\mathcal{D} \rangle$ state,
if inversion symmetry is kept. That is, the odd-$S$ Haldane phase is
a topological phase protected just by inversion symmetry. We emphasize that the
topological phase is characterized by the odd parity under inversion but
\emph{not} by a spontaneous symmetry breaking.

In contrast, an even-$S$ AKLT state
is even under inversion, regardless of the length of the
chain because there is an even number of
valence bonds on the links.  This argument suggests
that inversion does not protect the phase represented by
the even-$S$ AKLT state. Together with the previous section,
this supports our conjecture that the even-$S$ AKLT state is, in fact,
indistinct from a trivial state.


\subsection{Argument based on the Matrix-Product state representation}

The above heuristic argument, based on the global properties of the ground
state under inversion, requires the ring to have an odd length $L$. However,
this is not essential, as can be seen in the following more general
formulation~\cite%
{Pollmann-2010,Turner-2010}
based on matrix product states~\cite%
{FNW-MPS1992,Kluemper-spin1-JphysA1991} (MPS).
For completeness, we repeat the argument of Ref. %
\onlinecite{Pollmann-2010} below. Let us consider an inversion-symmetric
system. Although our analysis does not depend essentially on translation
symmetry, here we also assume a translation invariant MPS as in Eq.~%
\eqref{MPS}, for the sake of simplicity. On a chain of length $L$ with
periodic boundary conditions, a translation invariant MPS is given by
\begin{align}
|\Psi \rangle = & \sum_{m_1, \ldots, m_L} {\mathrm{Tr}} \big( A_{m_1} \ldots
A_{m_L} \big) | m_1, \ldots m_L \rangle,  \label{MPS}
\end{align}
where $A_{m}$ are $\chi \times \chi$ matrices, and $|m_j\rangle$
represents a local state at site $j$. We shall refer to the matrix
dimension $\chi$ as the ancilla dimension. We assume that the
ground state $|\Psi_0\rangle$ fulfills the following conditions:
\textbf{(a)} $|\Psi_0\rangle$ can be well approximated by the MPS
Eq. (\ref{MPS}) with finite dimensional matrices $A_m$,
\textbf{(b)} The matrices $A_m$ evolve continuously as we change a
parameter of the Hamiltonian, and \textbf{(c)} $|\Psi_0\rangle$ is
not a ``Cat state,'' i.e. a superposition of two states that are
not connected by any local operator (in analogy with
Schr\"odinger's cat or any superposition of two 
macroscopically different states).\cite{PerezGarcia-MPSrep}
The correlation length of an MPS state can be determined from
the eigenvalue spectrum of the completely positive map acting
on the space of $\chi \times \chi$ matrices\cite{PerezGarcia-MPSrep}
\begin{equation}
 \mathcal{E}(X) = \sum_m A_m X A^{\dagger}_m.
 \label{eq_CPM}
\end{equation}
This map can be interpreted as a transfer matrix which
determines correlation functions. The largest eigenvalue of
$\mathcal{E}$  for a normalized MPS is always equal to one. The
second largest (in terms of absolute value) eigenvalue
$\epsilon_2$ determines the largest correlation length
\begin{equation}
 \xi = - \frac{1}{\ln{|\epsilon_2|}} \label{eq:corr}
\end{equation}
for a state that is not a cat state.

It is useful to write the matrices $A_{m}$ in a canonical form as $A_{m}=\Gamma _{m}\Lambda $,
where $\Lambda$ is a diagonal matrix containing the square roots of the
eigenvalues of the reduced density matrix. The matrices $\Gamma _{m}$
and $\Lambda $ are then chosen to satisfy \cite{Vidal-2007,Orus-2008}
\begin{equation}
\sum_{m}\Gamma _{m}^{\dag }\Lambda ^{2}\Gamma^{\vphantom{\dag}} _{m}=%
\mathds{1}\text{ and }\sum_{m}\Gamma^{\vphantom{\dag}} _{m}\Lambda
^{2}\Gamma _{m}^{\dag }=\mathds{1}.  \label{canonical}
\end{equation}
This implies that the transfer matrix Eq.~(\ref{eq_CPM}) has an eigenvector $\mathds{1}$
with eigenvalue $\lambda =1$, and if condition (c) is fulfilled, all other
eigenvalues have smaller magnitudes.\cite{PerezGarcia-2008}

A reflection
corresponds to transposing all matrices $\Gamma _{m}\rightarrow \Gamma
_{m}^{T}$. This transformation preserves the canonical form of the MPS.
Since we assume the state to be invariant under inversion, we know from
Refs.~[\onlinecite{PerezGarcia-2008,Pollmann-2010}] that there exists a unitary $%
U_{\mathcal{I}}$ with $[U_{\mathcal{I}}, \Lambda]=0$ such that
\begin{equation}
\Gamma _{m}^{T}=e^{i\theta _{\mathcal{I}}}U_{\mathcal{I}}^{\dagger }\Gamma
_{m}U^{\vphantom{\dag}} _{\mathcal{I}}.  \label{eq:trans}
\end{equation}%
By iterating this relation twice, we arrive at $\Gamma _{m}=e^{2i\theta _{%
\mathcal{I}}}\left( U^{\vphantom{\dag}} _{\mathcal{I}}U_{\mathcal{I}}^{\ast
}\right) ^{\dagger }\Gamma _{m}U^{\vphantom{\dag}} _{\mathcal{I}}U_{\mathcal{%
I}}^{\ast }$. Combining this relation with Eq.~(\ref{canonical}), we obtain $%
\sum_{m}\Gamma _{m}^{\dag }\Lambda U^{\vphantom{\dag}} _{\mathcal{I}}U_{%
\mathcal{I}}^{\ast } \Lambda \Gamma^{\vphantom{\dag}} _{m}=e^{2i\theta _{%
\mathcal{I}}}U^{\vphantom{\dag}} _{\mathcal{I}} U_{\mathcal{I}}^{\ast }$.
I.e., the matrix $U^{\vphantom{\dag}}_{\mathcal{I}} U_{\mathcal{I}}^{\ast }$
is an eigenvector of the transfer matrix with an eigenvalue $e^{2i\theta_{%
\mathcal{I}}}$. Since we assume that all eigenvectors with unimodular
eigenvalues are proportional to $\mathds{1}$ with eigenvalue $\lambda=1$, $%
\theta_{\mathcal{I}} $ is either 0 or $\pi$ and $U^{\vphantom{\dag}}_{%
\mathcal{I}} U_{\mathcal{I}}^{\ast }=e^{-i\phi_\mathcal{I}}\mathds{1},$ or $%
U^T_{\mathcal{I}}=e^{i\phi_\mathcal{I}}U_{\mathcal{I}}$. Iterating the
latter relation twice, we find that $\phi_{\mathcal{I}}$ can be either $0$
or $\pi$, i.e., $U_\mathcal{I}$ is either symmetric or antisymmetric.
Eq.~(\ref{eq:trans}) implies that, for $\Gamma _{m}$ to evolve continuously,
$U_{\mathcal{I}}$ has to be continuous (up to a phase) and therefore must
remain symmetric or antisymmetric. Therefore, the only way in which $\phi _{%
\mathcal{I}}$ and $\theta _{\mathcal{I}}$ can change is through a phase
transition, in which one of the assumptions above break down. For example,
second order phase transitions through conformal critical points are
characterized by a diverging entanglement entropy and thus violate (a).\cite%
{Calabrese-2004} A violation of (b) corresponds to a first order
(discontinuous) phase transition. Violations of (c) correspond to a level
crossing in the spectrum of the transfer matrix $T$ which implies a quantum
phase transition, as discussed in detail in Ref.~\onlinecite{Wolf-2006}.

In the $S=1$ AKLT state, we can represent the state by a MPS with $%
\Gamma_{a}=\sigma_{a}/\sqrt{2}$. Here $\sigma _{a}$ ($a=x,y,z$) are Pauli
matrices and we use the time-reversal invariant spin basis $\left\vert
x\right\rangle =\frac{1}{\sqrt{2}}\left( \left\vert 1\right\rangle
-\left\vert -1\right\rangle \right) $, $\left\vert y\right\rangle =\frac{i}{%
\sqrt{2}}\left( \left\vert 1\right\rangle +\left\vert -1\right\rangle
\right) $, $\left\vert z\right\rangle =\left\vert 0\right\rangle $. Under
reflection of the system, the matrices transform as $\sigma _{a}\rightarrow
\sigma _{a}^{T}=-\sigma _{y}\sigma _{a}\sigma _{y}$. Therefore $U_{\mathcal{I%
}}=\sigma _{y}$ and $\theta_{\mathcal{I}}=\phi_{\mathcal{I}}=\pi$. We also
find $\theta_{\mathcal{I}}=\phi_{\mathcal{I}}=\pi$ for other odd $S$ while $%
\theta_{\mathcal{I}}=\phi_{\mathcal{I}}=0$ for even $S$. The state $|%
\mathcal{D}\rangle$, on the other hand, transforms simply as $%
\Gamma_m^T=\Gamma_m$ (since the $\Gamma_m$ are scalars) and thus $\theta_{%
\mathcal{I}}=\phi_{\mathcal{I}}=0$. Consequently, the system has to undergo
a phase transition when going from the odd-$S$ AKLT state to the trivial
state $|\mathcal{D} \rangle$, in agreement with the heuristic argument of
Sec. \ref{sec:inversion}.

\section{Triviality of the Even $S$ AKLT state}
\label{example}

We now complete the discussion by demonstrating that, on the other hand, the even-$S$ AKLT state is in the same phase as trivial states when
a number of different symmetries are imposed.
We begin by describing a state, formed by an MPS similar to
the one proposed in Ref. \onlinecite{Schuch-2010}, which shows that the $S=2$
AKLT state is smoothly connectable to the trivial state with
$S^z=0$ on every site. In this example, translation,
time-reversal, and inversion symmetries are maintained throughout
the path, but SU$(2)$ symmetry is broken. We continue by studying
an SU$(2)$ preserving example, in which we interpolate smoothly
from the $S=2$ AKLT state to a fully dimerized state. The same
construction fails for $S=1$, suggesting that in this case the
topological phase is protected in the presence of sufficient
symmetry, in agreements with our conclusion
that the odd-spin and even-spin states are distinct phases. 
Finally, we analyze a spin ladder example, in which an
interpolation from an $S=2$ AKLT state to a trivial state without
breaking any symmetry is possible.

\subsubsection{From $S=2$ AKLT to large-$D$ limit}

Let us present an explicit interpolation between the trivial state
$|\mathcal{D}\rangle$ (Eq. \ref{eq:Dstate}) and an even-$S$ AKLT
state, in terms of MPS. We focus on the $S=2$ case as the
simplest example. We take the standard $S^{z}$-basis so that $%
m_{j}=-2,-1,0,1,2$ can be identified with the eigenvalue of $S_{j}^{z}$. We
take $\chi =3$ as for the $S=2$ AKLT state and choose the matrices $A_{m}$
as a function of a parameter $t$:
\begin{equation}
A_{m}(t)\equiv tA_{m}^{\mbox{\scriptsize AKLT}}+(1-t)\delta
_{m,0}\left(
\begin{array}{ccc}
0 & 0 & 0 \\
0 & 1 & 0 \\
0 & 0 & 0%
\end{array}%
\right) .
\label{eq_interpolated}
\end{equation}%
Here, $A_{m}^{\mbox{\scriptsize AKLT}}$ is the MPS representation for the $%
S=2$ AKLT state~\cite{Totsuka-MPS1995}:
\begin{equation}
\sum_{m}A_{m}^{\mbox{\scriptsize AKLT}}|m\rangle =\frac{1}{\sqrt{10}}\left(
\begin{array}{ccc}
|0\rangle & \sqrt{3}|1\rangle & \sqrt{6}|2\rangle \\
-\sqrt{3}|-1\rangle & -2|0\rangle & -\sqrt{3}|1\rangle \\
\sqrt{6}|-2\rangle & \sqrt{3}|-1\rangle & |0\rangle%
\end{array}%
\right) .  \label{Aaklt}
\end{equation}%
By construction, the resulting MPS state $|\Psi (t)\rangle $ coincides with
the $S=2$ AKLT state at $t=1$, and reduces to the trivial state $|\mathcal{D}%
\rangle $ at $t=0$. $|\Psi (t)\rangle $ is not invariant under the global
SU(2) symmetry, except at $t=1$. On the other hand, it respects U(1)
symmetry (conservation of total $S^{z}$), translation symmetry, global D$%
_{2} $ symmetry, time-reversal symmetry, and inversion symmetry. Although $%
A_{m}(t)$ itself is not in the canonical form~\cite%
{PerezGarcia-MPSrep} of the MPS, for $0<t\leq 1$, it can be transformed to it.

Thus, if $|\epsilon_2|<1$ holds for $t$ in this range,
the correlation length remains finite and the path represents
an adiabatic evolution without a phase transition.
On the other hand, if $|\epsilon_2|=1$ occurs at a value of $t$
in $0 < t <1$, the correlation length diverges, signalling
a phase transition. 

\begin{figure}[tbp]
\includegraphics[width=8cm]{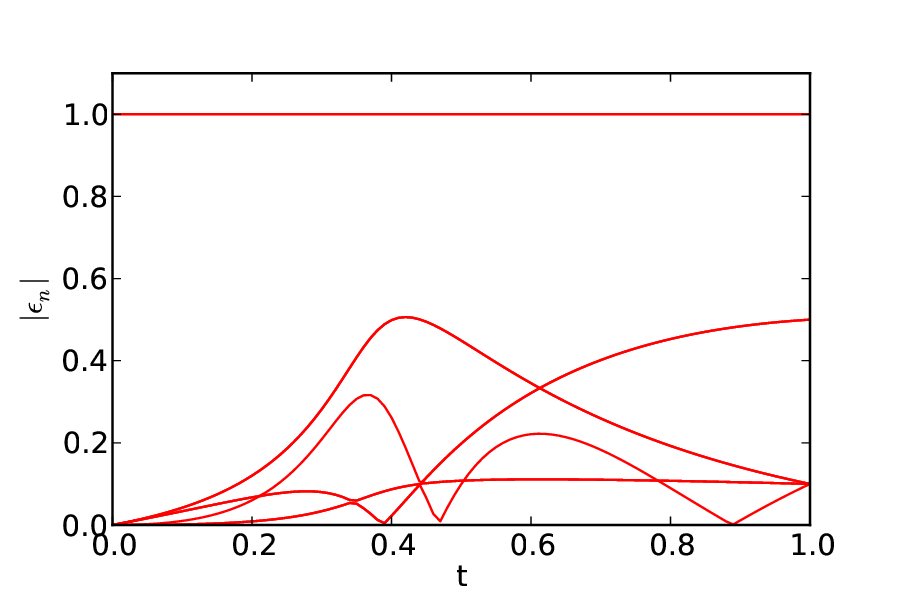}
\caption{(Color online) Eigenvalue spectrum of the transfer matrix
(completely positive map~\eqref{eq_CPM})
for the interpolating MPS defined in~\eqref{eq_interpolated}.
Except for the largest eigenvalue (unity), all the eigenvalues
have absolute value smaller than $1$ for $0 \leq t \leq 1$,
implying finite correlation length.
Thus the trivial state $|\mathcal{D}\rangle$ at $t=0$
and the $S=2$ AKLT state at $t=1$ are adiabatically connected
without any phase transition.}
\label{fig:Espectrum}
\end{figure}

The eigenvalue spectrum of $\mathcal{E}$ for the
interpolating MPS~\eqref{eq_interpolated} can be obtained
analytically as a function of $t$, using
\textsc{Mathematica}.
(The explicit expressions are lengthy and thus omitted here.)
The analytic expressions are plotted in Fig.~\ref{fig:Espectrum}.
The largest eigenvalue (without degeneracy) is unity
for all values of $t$, as expected.
Clearly, the absolute value of all the other eigenvalues are
smaller than $1$.
Thus the correlation length remains finite for $0 \leq t \leq 1$ and has no discontinuities,
implying that the $S=2$ AKLT
state is connected adiabatically to the trivial state $|\mathcal{D}\rangle $%
, without crossing any quantum phase transition. Moreover, the general
theorem of Ref.~\onlinecite{PerezGarcia-MPSrep} ensures that $|\Psi
(t)\rangle $ is the unique ground state of a Hamiltonian with only
short-range interactions and there is a nonvanishing excitation gap. Thus
the apparent nontrivial structure in the even-$S$ AKLT states is rather fragile;
these states can be be adiabatically connected to a trivial state
even while preserving inversion, D$_{2}$ and time-reversal symmetries.
This is in sharp contrast to the odd-$S$ case.

\subsubsection{From the AKLT state to a dimerized phase}

\begin{figure}[tbp]
\begin{tabular}{lc}
(a) & \includegraphics[width=7cm]{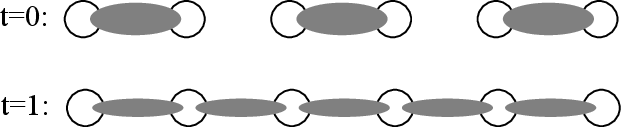}\\
(b) & \includegraphics[width=8cm]{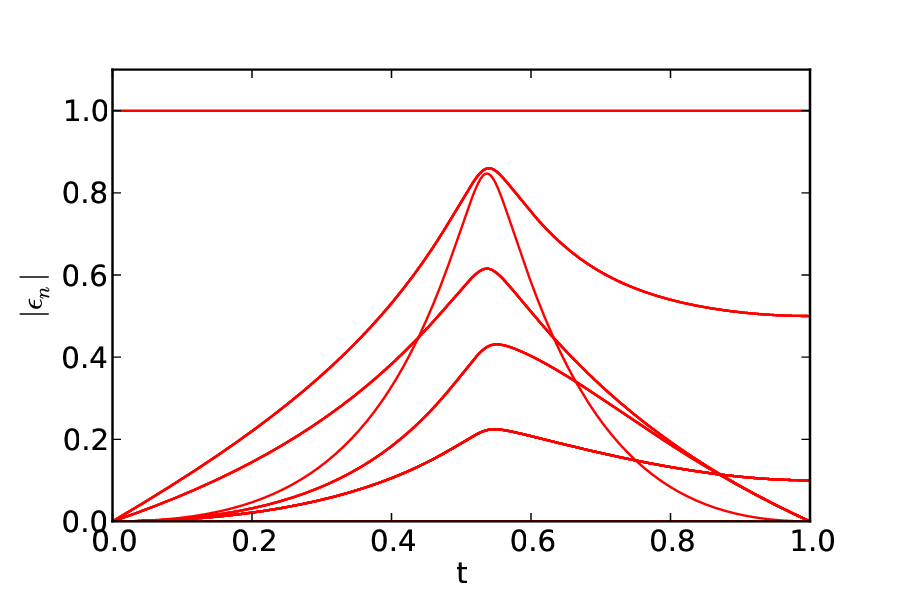}
\end{tabular}
\caption{(Color online) (a) Dimerized state of spin-$S$ singlets on every second
bond at $t=0$ and the AKLT state formed by $S/2$ singlets on every
bond at $t=1$. (b) Eigenvalue spectrum of the two-site transfer matrix
along the path connecting the fully dimerized state and the AKLT state for the spin $S=2$ chain (see text for details).}
\label{fig:dimer_aklt}
\end{figure}

In the particular example above we used an SU(2)-breaking path to connect
the $S=2$ AKLT state adiabatically to a trivial state. As explained in Sec. %
\ref{sec:edge},\ref{sec:inversion}, however, we expect that even if SU(2)
symmetry is respected, an even-$S$ AKLT state can be adiabatically connected
to a trivial state, if translational symmetry is broken. (Note that as long
as translational symmetry is retained, there is no path connecting the
AKLT state to a site-factorizable one, but for a trivial reason:
there is no site-factorizable state
with SU$(2)$ symmetry for $S>0$.)

We demonstrate this by constructing a continuous path in MPS\ space between
an $S=2$ AKLT\ state and a fully dimerized state. The explicit form of the
MPS\ $|\Psi \left( t\right) \rangle $ used to interpolate between the 
dimerized state at $t=0$
and a uniform AKLT\ state for $t=1$ is
given for general spin $S$ in Appendix \ref{AppedA} (see Fig.~\ref{fig:dimer_aklt}~(a)). The state  $|\Psi \left( t\right) \rangle $ is
invariant under SU$(2)$ and inversion for any value of $t$. To show that the
correlation length remains finite throughout the path, we have diagonalized
numerically the transfer matrix corresponding to $|\Psi _{t}\rangle $ (Eq. %
\ref{eq:trans}), for a range of values between $t=0$ and $1$. The results
for the 
are shown in
Fig.~\ref{fig:dimer_aklt}~(b). As can be seen in the figure,
although 
the correlation length is large, it remains finite for any
$0\le t \le 1 $. 

\subsubsection{Spin ladders}

\begin{figure}[tbp]
\begin{tabular}{lc}
(a) & \includegraphics[width=5cm]{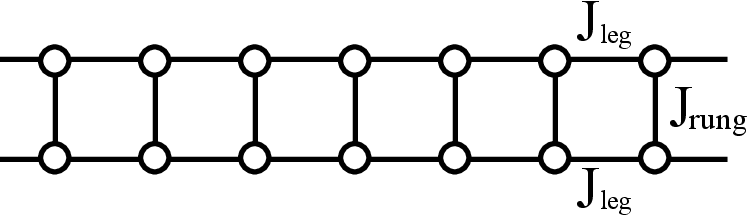} \\
(b) & \includegraphics[width=7cm]{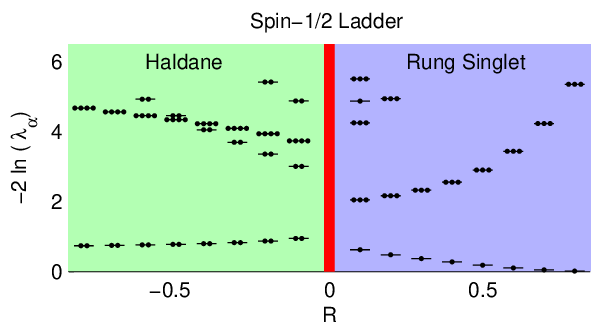} \\
(c) & \includegraphics[width=7cm]{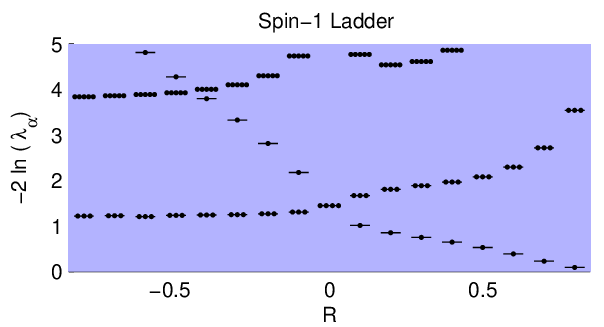} \\
&
\end{tabular}%
\caption{(Color online) (a) Ladder geometry used for the calculation. Entanglement spectra
for (b) $S=\frac12$ and (c) $S=1$ ladders. The entanglement spectrum is plotted
versus the ratio $R=J_{\text{rung}}/(J_{\text{leg}}+|J_{\text{rung}}|)$ for $%
J_{\text{leg}}>0$. Thus $R=-1$ corresponds to infinite ferromagnet coupling
on the rungs and $R=1$ to infinite antiferromagnet couplings. The number of dots on each level indicates its degeneracy.}
\label{fig:es}
\end{figure}

In order to illustrate the arguments in an especially intuitive way, 
we contrast $S=\frac12$ and $%
S=1$ two-leg ladder systems. The Hamiltonian is given by
\begin{eqnarray}
H &=&J_{\text{leg}}\sum_{i}\left\{ \mathbf{S}_{1,i}\cdot \mathbf{S}_{1,i+1}+%
\mathbf{S}_{2,i}\cdot \mathbf{S}_{2,i+1}\right\}   \notag \\
&+&J_{\text{rung}}\sum_{i}\mathbf{S}_{1,i}\cdot \mathbf{S}_{2,i},
\end{eqnarray}%
with the rung and leg couplings $J_{\text{rung}}$ and
$J_{\text{leg}}$, respectively. This Hamiltonian has been studied
extensively in the literature for both $S=\frac12$ and $S=1$ (for
example, see Refs.
\onlinecite{Dagotto-1996,Nishiyama-1995,Todo-2001, Poilblanc-2010, Laeuchli-2011}).
Here, we are interested in the case where we continuously tune the coupling $%
J_{\text{rung}}$ on the rungs from negative to positive values
while we keep the coupling on the legs constant 
($J_{\text{leg}}=1$). In the limit of $J_{\text{rung}}\rightarrow
-\infty $, the system maps to a $2S$ Heisenberg model. If
$J_{\text{rung}}\rightarrow \infty $, the spins form rung
singlets, and the ground states is a product state. In the $S=\frac12$
case, the point $J_{\text{rung}}=0$ is a critical point because
the systems corresponds to two decoupled $S=\frac12$ Heisenberg
chains, each having gapless excitations. Thus the two limits
cannot be connected adiabatically, at least through this path. In
the $S=1$ case, the point $J_{\text{rung}}=0$ corresponds to two
decoupled $S=1$ Heisenberg chains which are gapped. Furthermore,
previous Monte Carlo studies\cite{Todo-2001} have shown that the
correlation length remains finite along the entire path. Thus the
two limits are
connected adiabatically. This is in agreement with the above arguments: The $%
S=1$ Heisenberg model cannot be connected to a trivial product state while
the $S=2$ Heisenberg point can.

We used the infinite Time Evolving Block Decimation (iTEBD)\cite{Vidal-2007} algorithm to numerically
calculate the entanglement spectrum along the path connecting the two
limits. (see Fig.~\ref{fig:es}) 
The results are presented as a function of $R=J_{\text{rung}}/(J_{\text{leg}%
}+|J_{\text{rung}}|)$ for $J_{\text{leg}}>0$. In case of $S=\frac12$, we clearly
observe the predicted two-fold degeneracy in the Haldane phase.\cite%
{Pollmann-2010} At the critical point $R=0$, the entanglement spectrum
collapses to one point, and for positive $R$, the entanglement spectrum has
no double degeneracies any more.

In the $S=1$ case, the systems remains gapped along the entire
path and no divergence occurs, in agreement with the results of
Ref.~\onlinecite{Todo-2001}. Note that, in this case, the lowest
entanglement level is three-fold degenerate for $R<0$,
corresponding to an effective $S=1$ edge state, as expected in
an $S=2$ AKLT state. However, this edge degeneracy is not
sufficient to distinguish the AKLT state from the trivial (rung
singlet) phase. Indeed, at $R=0$, a level crossing occurs in the
entanglement spectrum, and for $R>0$ the lowest entanglement level
is singly degenerate, and the gap in the entanglement spectrum
increases monotonically with $R$, reaching that of the trivial
state for $R \rightarrow \infty$. Such a level crossing in the
entanglement spectrum can occur without a bulk phase transition.
In this sense, the existence of an edge state generally does {\em
not} define a phase.

\begin{figure}[tbp]
\includegraphics[width=7cm]{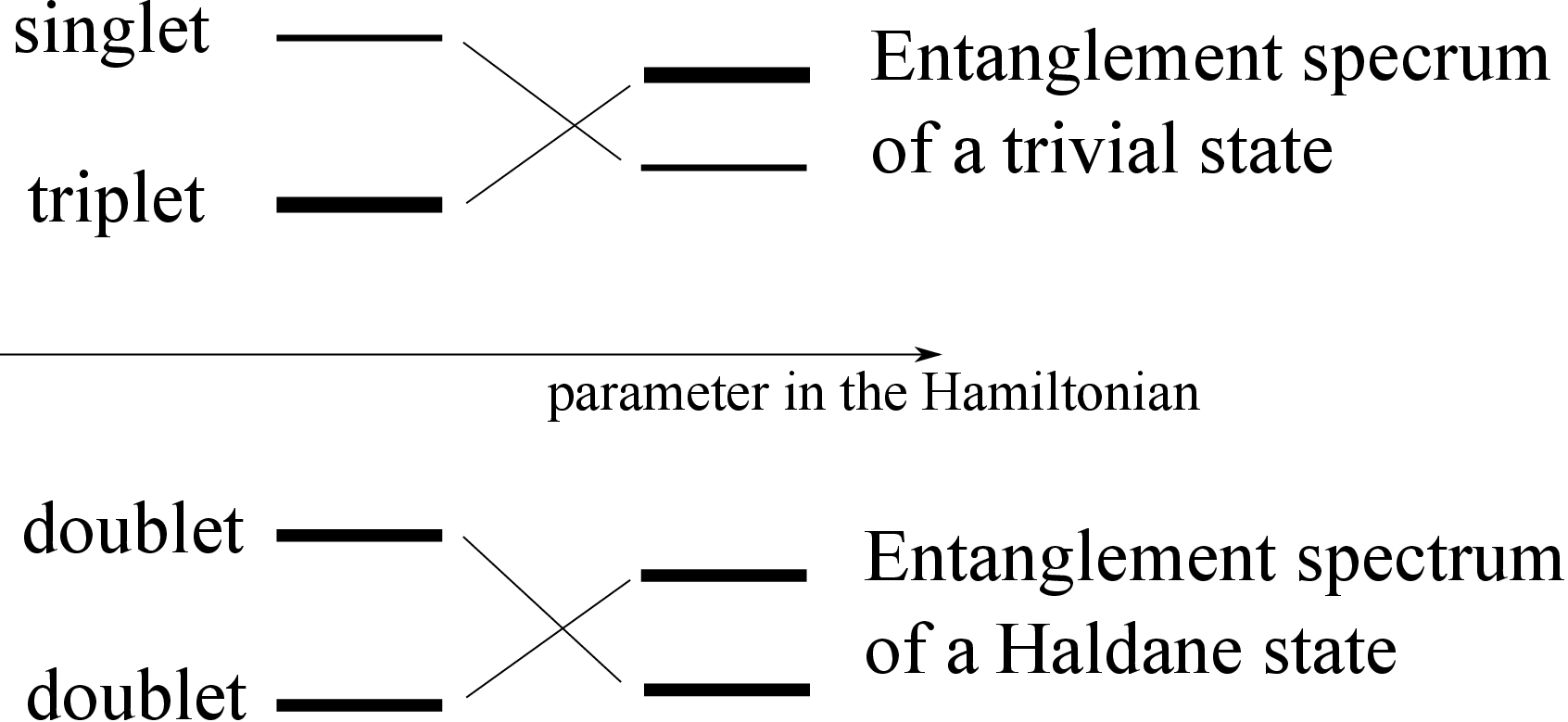}
\caption{Schematic diagrams of evolution of entanglement spectra.
In the trivial phase (upper panel), the degeneracy of the
lowest entanglement level can change, for example,
between unity and $3$ (which is indeed the case when $R$ changes
sign, in the $S=1$ two-leg lader discussed in the text).
This means that the edge state with $S=1$ can
appear and disappear without a bulk phase transition.
In contrast, in the Haldane phase (lower panel), the
entire entanglement spectrum is doubly degenerate.
Thus the lowest entanglement level is always doubly
degenerate, which implies existence of an edge state.
The edge state can be removed only via a bulk phase
transition (in the presence of an appropriate
symmetry described in the text).}
\label{fig:Levelcrossing}
\end{figure}

The difference between the two situations can be summarized as in
Fig.~\ref{fig:Levelcrossing}. In the Haldane phase, the entire
entanglement spectrum is at least doubly
degenerate.\cite{Pollmann-2010} Therefore, even if level
crossings occur, the lowest entanglement level is always doubly
degenerate. This degeneracy is related to an edge state, in the
presence of global D$_2$ or time-reversal symmetry. Such an edge
state is robust and signals a distinct phase. In the presence of
time-reversal symmetry, the protection of the edge state could
also be understood as a consequence of the Kramers degeneracy at
the edge; all the energy levels at the edge are doubly degenerate
and thus the edge state persists even in the presence of a level
crossing at the boundary. The degeneracy of the lowest
entanglement level or the edge state can be eliminated only via a
bulk phase transition (in the presence of an appropriate
symmetry), such as the critical point $R=0$ of the two-leg $S=\frac12$
ladder. This difference between robust and non-robust edge states
is also useful in understanding more complicated systems such as
spin tubes.\cite{Charrier-spintubes}




As a final remark, we comment on the relation of our results to those of
Anfuso and Rosch.\cite{Anfuso-2007} 
They constructed a path in the parameter space of a fermionic spin-$\frac12$
two-leg ladder model which adiabatically connects the $S=1$ Heisenberg point
(described as a Mott insulator with strong ferromagnetic interactions across
the rungs) with a trivial product state (a band insulator). The Hamiltonian
along this path is time-reversal and D$_2$ symmetric, in apparent
contradiction with our results for the odd-$S$ AKLT state. (Their model
breaks inversion symmetry explicitly along the path.) The reason for this
discrepancy is that the model of Anfuso et al. includes the possibility of
charge fluctuations; i.e., the elementary objects are not odd-$S$ spins, but
mobile $S=\frac12$ fermions. In that case, one cannot define uniquely the parity
of $\frac S2$ on a given site; even in the Mott insulator phase, virtual
fluctuations in the fermion number can switch the site from integer to
half-integer $S$. Therefore, the arguments presented above for protection by
time-reversal or D$_2$ symmetries, which relied crucially on the fact that
every site has 
a well-defined spin, break down. In contrast, in the models we
consider here, we assume that the particles are immobile, and
therefore the local spin is well-defined. On the other hand, in
the presence of a lattice inversion symmetry, we expect that the
Haldane phase is still robust as a topological phase, even in
fermionic models. This does not contradict with
Ref.~\onlinecite{Anfuso-2007}, as their model breaks the inversion
symmetry explicitly.

\section{Conclusions}

To summarize, we have shown that the topological phase in the odd-$S$ AKLT
state is protected as long as either time-reversal, link-centered inversion,
or global D$_2$ rotational symmetry is preserved.
This symmetry protection,
which has been argued before\cite{Pollmann-2010} on the basis of properties
of the entanglement states, is shown to follow from simple physical
arguments: the D$_2$ protection is a result of the hidden
symmetry breaking in the AKLT state\cite{KennedyTasaki-PRB1992}; protection
by time-reversal symmetry is related to the Kramers degeneracy due to
the effective half-integer edge spin in an odd-$S$ AKLT state; and the
protection by link inversion follows from the odd parity of this state under
inversion. Note that none of the above arguments depends on translational
symmetry.

In contrast, we argue that even-$S$ AKLT states are fundamentally different.
Even in the presence of all the above-mentioned symmetries, this state is
adiabatically connectable to a trivial state, as we demonstrated explicitly
using a path in the MPS space.
A similar adiabatic connection between the $S=2$ AKLT state and
a trivial state is suggested recently in the
analysis of finite-length $S=2$ chains with both exchange
and single-ion anisotropy.\cite{Tonegawa-Spin2}
Even if the full SU(2)
invariance is maintained along the path, we have demonstrated that
an even-$S$ AKLT state is adiabatically connectable to a trivial
dimerized state with broken translational symmetry.



Our analysis can be extended to more general one dimensional quantum spin
systems, such as chains with bond alternation, spin ladders and tubes. In
the AKLT-type construction based on valence bonds, when a ``cut'' (such as
the vertical line in Fig.~\ref{fig:AKLTringparity}) is crossed by an odd
number of valence bonds, the state has a robust topological phase protected
by either time-reversal or link-inversion symmetry, thanks respectively to
the edge Kramers degeneracy or the odd parity with respect to link
inversion. For example, the $n$-leg $S=1$ Heisenberg ladder, in the weak
rung coupling limit (where each chain becomes independent), is
topologically distinct from a product state when $n$ is odd. The topological phase survives at a finite rung coupling (persisting until
the system passes through a quantum transition) provided that
either time-reversal or link inversion is kept.


\acknowledgements

We thank S. Capponi, D. Charrier, J.~E. Moore, K. Okamoto,
K. Okunishi, P. Pujol, and T. Tonegawa for useful
discussions. This work was supported by ARO grant W911NF-07-1-0576 (F.~P.
and A.~M.~T.), by NSF grants DMR-0757145 and DMR-0705472 (E.~B.), and by
KAKENHI grants 20654030 and 20102008 from MEXT of Japan (M.~O.). F.~P. and
M.~O. acknowledge the Workshop 
\textit{TOPO09} at MPIPKS Dresden, where a part of the present work was
carried out.

\appendix

\section{MPS path from a dimerized state to the AKLT state}

In this appendix, we construct a path between a fully dimerized
state (a broken translational symmetry state in which every pair
of spins are coupled in a singlet) and an AKLT state, for general
spin $S$, which remains gapped when the spin is even. 
The path is parametrized by a parameter $t$, where $t=0$
corresponds to the fully dimerized state and $t=1$ corresponds to
the AKLT\ state. The MPS along the path is of the form
\begin{eqnarray}
|\Psi \rangle =  \sum_{m_1, \ldots, m_L} {\mathrm{Tr}} &\big(& A_{m_1}(t) B_{m_2}(t) \ldots
A_{m_{L-1}}(t)B_{m_L}(t) \big) \times\nonumber\\
&|& m_1, \ldots m_L \rangle,
\end{eqnarray}where the  matrices $A _{m}(t)$, $B_{m}(t)$ are given by:
\begin{eqnarray*}
A _{m}\left( t\right) =\left(
\begin{array}{cc}
tA^{11}_{m} & {t(1-t)}A^{12}_{m} \\
{t(1-t)}A^{21}_{m} & (1-t)A^{22}_{m}%
\end{array}%
\right) \text{,}\nonumber\\
B_{m}\left( t\right) =\left(
\begin{array}{cc}
tB^{11}_{m} & {t(1-t)}B^{12}_{m} \\
{t(1-t)}B^{21}_{m} & (1-t)B^{22}_{m}%
\end{array}%
\right) \text{.}
\end{eqnarray*}%
Here, $A^{11}_{m}$ is an $\left( S+1\right) \times \left( S+1\right) $ matrix, $%
A^{12}_{m}$ is $\left( S+1\right) \times 1$, $A_{m}^{21}$ is
$\left( 2S+1\right) \times \left( S+1\right) $, and $A_{m}^{22}$
is $\left( 2S+1\right) \times 1$. The dimensions of $B^{ij}_{m}$
are the same as those of $(A^{ji}_{m})^{T}$.


As $t$ varies from $1$ to $0$, the state evolves from the AKLT state,
which corresponds to the upper left blocks of the $A$ and $B$
matrices, into the dimerized state, which is defined by the lower
right block of $A$, $B$.
For intermediate values of $t$, the off-diagonal blocks $A^{12}$,
$A^{21}$, $B^{12}$, $B^{21}$ mix these two states together.

The matrix elements of the matrices
$A^{11}_{m},B^{11}_{m}$ are given by
\begin{equation}
\left[ A^{11}_{m}\right] _{\alpha,\beta}=[B^{11}_{m}]_{\alpha,\beta}=\left(
-1\right) ^{\beta}\langle S/2,\alpha,S/2,\beta;S,m\rangle \text{,}
\end{equation}%
where $\langle j_{1},m_{1},j_{2},m_{2};J,M\rangle $ are
Clebsch-Gordan coefficients, and we index the matrix elements by
$\alpha$,$\beta=-S/2,\dots ,S/2$.  The matrices $A^{11}_m$, $B^{11}_m$ are exactly the matrices of the AKLT state in Eq.~(\ref{Aaklt}).

Similarly, the matrix elements of
the other matrices are
\begin{equation}
\left[ A^{12}_{m}\right] _{\alpha,\beta}=[B^{21}_{m}]
_{\alpha,\beta}=0,
\end{equation}

\begin{eqnarray}
\left[A^{21}\right] _{\alpha,\beta} &=&\left( -1\right) ^{\beta}\langle
S,\alpha,S/2,\beta;S,m\rangle ,  \notag \\
\lbrack B^{12}_{m}]_{\alpha,\beta} &=&\left( -1\right) ^{\beta}\langle
S/2,\alpha,S,\beta;S,m\rangle ,
\end{eqnarray}

\begin{eqnarray}
\left[A^{22}\right] _{\alpha,\beta} &=&\left( -1\right) ^{\beta}\langle
S,\alpha,0,\beta;S,m\rangle ,  \notag \\
\lbrack B^{22}_{m}]_{\alpha,\beta} &=&\left( -1\right) ^{\beta}\langle
0,\alpha,S,\beta;S,m\rangle.
\end{eqnarray}\\

Now for odd $S$, this path fails (as we expect) to be continuous.
The matrix $A^{21}$ vanishes because it is not possible to make a
spin $S$ state (which is an integer) out of integer- and
half-integer spin particles $S$ and $S/2$. Thus, the AKLT state
and the dimerized state are not combined with one another in any
way, and the transition is discontinuous. The state is just
$t^N|\text{AKLT}\rangle+(1-t)^N|\text{Dimerized}\rangle$; at a
certain value of $t=t_c$, the two terms have equal weights.
Everywhere else, one of the two terms is exponentially bigger than
the other, so the correlation functions change discontinuously
from the AKLT state's correlation functions to the dimerized
state's correlation function.
 (The correlation length of the MPS
cannot be calculated using Eq.~\ref{eq:corr} in this case because the state is not a pure states in the sense of Ref.~\onlinecite{Wolf-2006})


\label{AppedA}


\begin{thebibliography}{40}
\bibitem[{com()}]{comment-TR}
\bibinfo{note}{In Ref.~\onlinecite{GuWen-TEFR2009}, the term ``time reversal''
  is used for complex conjugation in the $S^z$-basis, namely $S^{x,z}_j
  \rightarrow S^{x,z}_j , S^y_j \rightarrow - S^y_j$. However, in this paper,
  we refer to the transformation $S^{x,y,z}_j \rightarrow - S^{x,y,z}_j$ as
  time reversal.}
\end{thebibliography}

\end{document}